\documentclass[prb, twocolumn,superscriptaddress,preprintnumbers,a4paper,amsmath,amssymb,showpacs,floatfix]{revtex4}

\usepackage{graphicx}
\usepackage{color}\color[rgb]{0.000,0.000,0.000} 
\usepackage{amssymb} 
\usepackage{amsmath} 

\begin{document}

\newcommand{\pr}{PrFeAsO}
\newcommand{\prf}{PrFeAsO}
\newcommand{\prsc}{PrFeAsO$_{0.85}$F$_{0.15}$}

\newcommand{\te}{$_{t}$}
\newcommand{\ot}{$_{o}$}

\newcommand{\Tc}{T$_{C}$}
\newcommand{\Ts}{T$_{s}$}
\newcommand{\Tn}{T$_{N}$}
\newcommand{\Tnpr}{T$_{N}$(Pr)}
\newcommand{\Tnfe}{T$_{N}$(Fe)}
\newcommand{\Tnferho}{T$_{N}^{\rho}$(Fe)}


\title{Magnetic Ordering and Negative Thermal Expansion in \prf}

\author{S. A. J. Kimber}
\affiliation{Helmholtz-Zentrum Berlin f\"ur Materialien und Energie (HZB), Glienicker Strasse 100, D-14109, Berlin, Germany}

\author{D.~N.~Argyriou}
\email[Email of corresponding author:]{argyriou@helmholtz-berlin.de}
\affiliation{Helmholtz-Zentrum Berlin f\"ur Materialien und Energie (HZB), Glienicker Strasse 100, D-14109, Berlin, Germany}

\author{F. Yokaichiya}
\affiliation{Helmholtz-Zentrum Berlin f\"ur Materialien und Energie (HZB), Glienicker Strasse 100, D-14109, Berlin, Germany}

\author{K. Habicht}
\affiliation{Helmholtz-Zentrum Berlin f\"ur Materialien und Energie (HZB), Glienicker Strasse 100, D-14109, Berlin, Germany}

\author{S. Gerischer}
\affiliation{Helmholtz-Zentrum Berlin f\"ur Materialien und Energie (HZB), Glienicker Strasse 100, D-14109, Berlin, Germany}

\author{T. Hansen}
\affiliation{Institute Max von Laue-Paul Langevin, 6 rue Jules Horowitz, BP 156, F-38042, Grenoble Cedex 9, France}

\author{T. Chatterji}
\affiliation{JCNS, Forschungszentrum J\"ulich Outstation at Institut Laue-Langevin, BP 156, F-38042, Grenoble Cedex 9, France}

\author{R. Klingeler}
\affiliation{Leibniz-Institute for Solid State and Materials Research (IFW) Dresden, Germany}

\author{C. Hess}
\affiliation{Leibniz-Institute for Solid State and Materials Research (IFW) Dresden, Germany}

\author{G. Behr}
\affiliation{Leibniz-Institute for Solid State and Materials Research (IFW) Dresden, Germany}


\author{A. Kondrat}
\affiliation{Leibniz-Institute for Solid State and Materials Research (IFW) Dresden, Germany}

\author{B. B\"uchner}
\affiliation{Leibniz-Institute for Solid State and Materials Research (IFW) Dresden, Germany}

\date{\today}

\begin{abstract}
We report the structure and magnetism of PrOFeAs, one of the parent phases of the newly discovered Fe-As superconductors, as measured by neutron powder diffraction. In common with other REOFeAs materials, a tetragonal-orthorhombic phase transition is found on cooling below 136 K and striped Fe magnetism with $k =$(1,0,1)  is detected below $\sim$ 85 K.  Our magnetic order parameter measurements show that the ordered Fe moment along the $a$ axis reaches a maximum at $\sim$ 40 K, below which an anomalous expansion of the $c$ axis sets in, which results in a negative thermal volume expansion of 0.015 \% at 2 K. We propose that this effect, which is suppressed in superconducting samples, is driven by a delicate interplay between Fe and Pr ordered moments.
\end{abstract}

\maketitle


Until very recently, and despite decades of research, no new high Tc superconductors had been
found- perhaps due to the perceived wisdom that the only likely candidates were doped Mott
insulators. In this context, it is therefore difficult to overestimate the importance of the
discovery of superconductivity at 25 K in $LaFeAsO_{1-x}F_{x}$ \cite{Kamihara:2008p6110}. Like the
cuprates, the $RE$FeAsO ($RE$=rare earth) family of compounds are layered, however, the parent
phase is a bad metal and iron is in a tetrahedral coordination with arsenic in contrast to the
square planar geometry of the cuprates. On cooling, the undoped parent phases undergo an
electronic transition at \Ts$\sim$150 K at which the resistivity and magnetisation drop sharply.
Initial powder diffraction studies showed that this transition corresponds to a structural change
from tetragonal to orthorhombic symmetry \cite{delaCruz:2008p6840}, and at slightly lower
temperatures, commensurate Fe-spin order
develops\cite{Nomura:2008p6126,Cruz:2008p6042,Klauss:2008p6055}.

Initially, many authors speculated that the structural distortion is the result of a spin density
wave transition and that superconductivity arises as a result of doping which suppresses both the
structural phase transition and the Fe-spin ordering \cite{Dong:2008p6037}. On this basis, strong
analogies have been drawn with the cuprate materials. Alternatively, as the Fe-As layers in the
high temperature tetragonal phase of these compounds are a perfect realization of the
$J_{1}-J_{2}$ frustrated square lattice, magnetic frustration has also been proposed as a
potential driving force for the symmetry lowering phase transition at 150 K
\cite{Yildirim:2008p6916}. Indeed, the magnetic structures reported for the majority of the
undoped compounds are simple collinear striped models, which are predicted to be the ground state
of the $J_{1}-J_{2}$ model when $J_{1}/J_{2}\sim$1 [7].  Evidence for coupling between structural
and magnetic order parameters is particularly strong in the $AFe_{2}As_{2}$ (A = Ca, Sr, Ba)
family of materials. Neutron diffraction studies show that magnetic order emerges exactly at the
structural transition \cite{Huang:2008p6917,Zhao:2008p6783,Goldman:2008p6925}. In the $RE$FeAsO
compounds however, the structural and magnetic transitions are well separated, in LaFeAsO
\cite{delaCruz:2008p6840} and CeFeAsO \cite{Zhao:2008p6522}, spin order is found some 20 K below
\Ts\ suggesting a weaker coupling.

In this communication we describe a series of neutron scattering experiments which show that in
\pr\ there is a magneto-elastic coupling that results in a negative thermal expansion (NTE) of the
$c-$axis below $\sim$40 K and an expansion of the unit cell volume.  While for this compound
\Ts=136 K, our measurements become sensitive to Fe magnetic order much below \Ts\ at a tentative
\Tnfe=85 K, the Fe spins order in the form of antiferromagnetically coupled stripes with the
moments pointing along the $a-$axis and propagation vector $k=$(1,0,1). The Pr spins are found to
order below \Tnpr=12 K.  The emergence of the lattice anomaly correlates with changes in the
intensity of the $(1\overline{1}2)$ magnetic reflection and is likely driven by a subtle interplay
between Pr and Fe spin order. Our diffraction measurements show that the NTE behavior is
suppressed in superconducting \prsc\ together with magnetic order.

We synthesised polycrystalline samples of \pr\ and \prsc\ by the previously reported
methods\cite{Zhu2008p6810}. The resistivity data for PrFeAsO shown in Fig.~\ref{fig1} exhibit
several anomalies below 300 K. The most prominent feature is a broad maximum around \Ts.  While
the resistivity increases when approaching \Ts\ from higher temperatures there is a pronounced
suppression of $\rho$ around 150 K. Such a behavior is typical for the REFeAsO materials.
Interestingly, similar to e.g. LaFeAsO, there is a kink in the derivative d$\rho$/d$T$ at a
temperature  slightly below \Ts .\cite{Klauss:2008p6055} From the comparison with LaFeAsO we
conclude  that this kink indicates the onset of long range spin order as indirectly probed by the
electrical resistivity, i.e \Tnferho\ $\sim$127 K. Note, that in contrast to LaFeAsO large RE
moments are present in PrFeAsO which mask any anomaly in our static magnetisation data around \Ts.
An additional anomaly, i.e.  a sharp decrease of the resistivity, is observed at much lower
temperatures, i.e. \Tnpr=11K. Our neutron data (see below) confirm that at this temperature
magnetic order of the Pr spins evolves. Our fluorine doped sample shows a superconducting
transition with a \Tc=43 K.  Neutron powder diffraction  (NPD) measurements were measured as a
function of temperature on the D20 powder diffractometer\cite{Hansen:2008p034001} located at the
Institut Lau-Langevin using a wavelength of $\lambda$=1.88 \AA. For these measurements we
collected NPD data continuously every 60 seconds while the sample was warmed at a rate of 2 K/min.
between 1.4 and 100 K and 1 K/min between 100 and 160 K. Measurements on cooling between 200 and
150 K were made in a similar fashion with a cooling rate of 2 K/min. This mode of operation allows
for a high degree of flexibility in the data collection and analysis. While the high intensity of
D20 allowed us to collect high quality NPD data within a 1 minute period suitable for Rietveld
refinement, when better statistics are desired the data can be  integrated over a desired
temperature range (typically over a few K). Additional NPD data were also  measured using the E9
high resolution neutron powder diffractometer with $\lambda$=1.79 \AA\ and the FLEX cold triple
axis spectrometer with $\lambda$=4.05 \AA, both at HZB. The NPD data were analysed using the
Rietveld method with the programs GSAS\cite{Larson:1998p6951} and FULLPROF
\cite{Roisnel:2001p6953}.

\begin{figure}[tb!]
\begin{center}
\caption{(color onine) (a) Temperature evolution of the PrOFeAs (220)\te\ peak as measured at
$2\theta$=82.3 deg. and FWHM of the (220)\te\ and (400)/(040)\ot\ peaks. (b) Resistivity
measurements from the \pr\ (c) and superconducting \prsc\ sample. }
\label{fig1}
\end{center}
\end{figure}

Turning our attention on the NPD measurements from D20, we find that on cooling the PrOFeAs sample, the structural phase transition from P$4/nmm$ to C$mma$ occurs at \Ts=136 K, similar to what has been reported for the other REOFeAs materials \cite{delaCruz:2008p6840,Bos:2008p6802,Qiu:2008p6803,Nomura:2008p6126}. The evolution with temperature of the (220)\te\ tetragonal reflection, which splits on cooling, into the (400)\ot\ and (040)\ot\ reflections is shown in Fig.~\ref{fig2}(a). The sharp correlation between the resisitivity decrease at 136 K and the structural phase transition is highlighted by plotting the intensity of the (220)\te\ peak from the NPD data between 83.3 and 83.4 deg as a function of temperature (fig.~\ref{fig1}).  The exact determination of the transition temperature from the NPD data is challenging as the nature of this transition appears to be first order\cite{delaCruz:2008p6840,Goldman:2008p6925}. The exact transition temperature can not be established simply by the sequential Rietveld refinements of the NPD data as the data can be almost equally well fitted just above or below \Ts\ with either the C$mma$ or P$4/nmm$ structures.  A better method to analyse the splitting of the (220)\te\ reflection close to \Ts\ is to model it with either one or two resolution limited reflections.  This defines the transition temperature (for example on cooling) as the point of which the (220)\te\ reflection becomes significantly broader that the resolution of the diffractometer to support modelling by two peaks.  Such analysis shows that the full width at half maximum (FWHM) of the (400)/(040)\ot\ reflection sharply increases at \Ts =136 K in excellent agreement with that obtained from the resisitivity measurements, while the (220)\te\ reflection is signficantly broader above \Ts. A physical interpretation of such analysis would suggest an abrupt transition on warming with an extended range of co-existance as a broader peak within the resolution of the measurement. 

\begin{figure}[tb!]
\begin{center}
\caption{(color online)(a) Temperature evolution of the (220)\te\ peak in PrFeAsO (D20 data)
showing the orthorhombic splitting at \Ts. (b) Temperature dependence of the $a-$ and $b-$axis
(top) and $c- $axis (bottom) determined from Rietveld analysis of the neutron powder diffraction
data. A clear anomalous expansion of the $c-$axis, which corresponds to the layer stacking
direction is observed below ~50 K. This expansion is confirmed by measurement on the E9
diffractometer shown on the same panel as (blue) open symbols.} \label{fig2}
\end{center}
\end{figure}

The effect of the structural distortion on the cell parameters is also shown in Fig. ~\ref{fig2}(b). The $a-$axis is compressed and the $b-$axis expands leading to inequivalent $J_{[100]}$ and $J_{[010]}$ interactions in the Fe-As planes as the Fe-Fe distances and Fe-As-Fe bond angles change. These structural changes lift the magnetic frustration,  allowing the development of Fe spin order \emph{(vide infra)} at lower temperature. Surpisingly on further cooling below $\sim$ 45 K the $c-$axis (which corresponds to the layer stacking direction) smoothly expands. As shown in Fig.~\ref{fig4}, this effect dominates the unit cell volume at low temperature, leading to a small, but measurable, negative thermal expansion of the volume (see below). This increase of the $c-$axis was confirmed at low temperatures also on the E9 diffractometer at HZB in a constant temperature mode of operation  (shown as black circles in Fig.~\ref{fig2}(c)), confirming our measurements on D20.  In order to estimate the size of the effect, we fitted the high temperature (50 Ð 160 K) cell volume to a 2nd order Gr\"uneisen approximation yielding a Debye temperature $\theta_{D}$ = 355(5) K \cite{Vocadlo:2002p7169}. The extrapolated cell volume at low temperature is shown as a solid line in Fig.~\ref{fig4}. Comparing the expected unit cell volume on the basis of the 2nd order Gr\"uneisen fitting we calculate a volume change  of  ca. 0.015 \% at 1.4 K, which is remarkably similar in magnitude to the magnetoelastic volume  effect and negative thermal expansion reported in other frustrated magnets, such as the spinels $ZnCr_{2}Ch_{4}$ (Ch = S, Se) \cite{Hemberger:2007p6753}. We propose that negative thermal expansion in PrFeAsO originates in magnetoelastic coupling related to the onset of Fe and Pr spin order. Below \Ts\ we find evidence of magnetic ordering arising from Pr and Fe spins.  On cooling below \Ts\ we observe the emergence of magnetic reflections, the most prominent being the  (101) and (102). These reflections index on the chemical cell and correspond to a simple striped model of Fe spins with moments of 0.35(5) $\mu_{B}$ along the $a-$axis with $k =$(1,0,1)  at 30 K as shown in fig.~\ref{fig3}. At 12 K we also find that Pr-spins order giving a sharp rise in  the intensity of these magnetic reflections as well as the observation of a new  (100) magnetic reflection. The best fit to the low temperature NPD data was achieved by a model where both Fe- and Pr-spins order within the $ab-$plane in the two-left, one-right magnetic structure described for NdOFeAs \cite{Qiu:2008p6803,Bos:2008p6802,Chen:2008p6809}. At 1.4K the moments of the Fe- and Pr- ions are 0.53(20) and 0.83(9) $\mu_{B}$ respectively. We searched extensively, but unsuccessfully for reflections which would justify a change in the magnetic propagation vector below 12 K, as was reported for $RE$=Ce \cite{Chen:2008p6809}. In addition we attempted to fit the magnetic scattering at low temperature with the Fe spin model described above and various arrangements of Pr moments, with and without doubling the $c-$axis. The model described above provided the best fit to the data by far.

\begin{figure}[tb!]
\begin{center}
\caption{(a) Magnetic order parameter for PrOFeAs measured at the (102)\ot reflection; (b)
Magnetic order parameter measured at the (101) reflection; (c) Refined cell volume of PrFeAsO from
neutron powder diffraction data, red circles represent data collected using D20 at the ILL, blue
circles are data collected using E9 at HZB. Also shown is a Debye fit to the high temperature (50
- 120 K) cell volume.} \label{fig4}
\end{center}
\end{figure}

The onset of the Pr and Fe spin ordering was determined by following the temperature dependence of
the magnetic reflections using both the D20 data as well as performing additional measurements on
the FLEX spectrometer. The temperature dependence of the (101) and (102) magnetic reflections are
shown in fig. \ref{fig4}(a) and (b).  Given the small size of the ordered Fe moment, the D20 data
were binned in approximately 5 K intervals to enhance the statistics of a given diffraction
pattern. To obtain the temperature dependence of the (102) reflection the data were integrated
over a 1.5 deg range around the  peak.  A similar size range of $2\theta$ adjacent to the (102)
reflection was also integrated to gain an estimate of the background and its temperature
dependence, which was found to be constant.  The results of this analysis are plotted in
fig.~\ref{fig4}(a).  We find that although there is a well defined Bragg reflection at low
temperature the magnetic scattering becomes much weaker as temperature is increases towards \Ts.
A sharp drop in the intensity of this reflection is observed on warming through 12 K, which
signifies the onset of Pr ordering, and the magnetic scattering  finally disappears above 85 K.
This observation will tentatively place \Tnfe$\sim$85 K,  however, In the present investigation we
are only sensitive to magnetic scattering once the Fe moment reaches a value of ~0.1 $\mu_{B}$.
more sensitive probes to Fe-magnetism such as muon and M\"ossbauer spectroscopy may well place
\Tnfe\ to be either coincident with \Ts\ or are a few Kelvin below as it has been found for
$RE$=La, Ce and Nd \cite{Carlo:2008p6043,delaCruz:2008p6840,Zhao:2008p6522}.

We followed the temperature dependence of the (101) reflection with the FLEX triple axis
spectrometer and found its evolution to be somewhat different from that of the (102).  A large
decrease is also observed for this reflection on warming through  \Tnpr\ , however, the scattering
from the Fe spin order reaches a plateau around 40 K before disappearing above 75 K. The apparent
lower \Tnfe\ given by this reflection likely arises from its smaller magnetic structure factor
compared to the (102).

\begin{figure}[tb!]
\begin{center}
\caption{(left) Low angle region of neutron powder diffraction patterns for PrFeAsO, recorded
using D20. Top pattern shows fit at 30 K in Fe spin ordered regime, bottom pattern shows combined
Fe and Pr fit at 1.4 K. Stars mark areas of disagreement between observed and calculated patterns;
(right) Refined Fe magnetic structure of PrFeAsO at 30 K.} \label{fig3}
\end{center}
\end{figure}

As is evident in Fig.~\ref{fig4} the expansion of the unit cell volume coincides with the plateau
of the (101) reflection suggesting that its origin is magnetoelastic. The rise in intensity of the
(101) Fe magnetic reflection up to $\sim$40 K and its decrease as \Tnpr\ is approached suggest a
possible interplay between Pr and Fe. One possible explanation is a decrease in the ordered Fe
moment along the $a$ axis due to the influence of the Pr moments.  Such behavior has also been
alluded to in ref. \cite{Zhao:2008p6783} for CeFeAsO, however in the present case the
magneto-elastic coupling that is generated is sufficiently strong to produce an increase of the
unit cell volume.  Such effects have not been observed to date in the $REFeAsO$ family of
compounds.

We also performed neutron powder diffraction measurements on superconducting \prsc\ down to 0.4 K
using a $^{3}He$ insert. We detect no evidence for Fe or Pr magnetic order, while the NPD data
show a positive thermal expansion for the $a-$ and the $c-$axis. This result is also in contrast
to measurements for cuprate superconductors such as Nd$_{2-x}$Ce$_{x}$CuO$_{4}$, in which rare
earth antiferromagnetism is found to co-exist with the superconducting state\cite{Lynn:1990p2569}
below ~ 2.5 K. This would indicate that the rare earth magnetism in the layered FeAs compounds may
be induced by the Fe-spin order.

In conclusion, our measurements show that PrFeAsO undergoes the same T - O phase transition as the
other REFeAsO compounds at 136 K. On cooling below $\sim$85 K we detect long range magnetic order
of the Fe moments in a striped arrangement, with $k =$(1,0,1) and at lower temperatures (12 K)
long range order of Pr spins. Our detailed Rietveld refinements have identified a previously
unknown coupling of magnetic and lattice degrees of freedom at  $\sim$40 K, which leads to an
anomalous expansion of the layer stacking direction. We propose that that this negative thermal
expansion results from a delicate interplay between the ordered Fe and Pr sublattices. Suppression
of the NTE state and long range magnetic order is necessary for the emergence of superconductivity
in  Pr$_{1-x}$F$_{x}$FeAsO.

We acknowledge the Helmholtz Zentrum Berlin for funding and the Institute Max von Laue-Paul
Langevin for access to their instruments.



\end{document}